\documentclass[12pt]{iopart}
\usepackage{graphicx}
\usepackage{color}
\usepackage{dcolumn}
\usepackage{bm}
\newcommand{\vpsi}{\boldsymbol{\psi}}

\newcommand{\vx}{\boldsymbol{x}}
\newcommand{\vw}{\boldsymbol{w}}
\newcommand{\vtheta}{\boldsymbol{\theta}}
\newcommand{\vI}{\boldsymbol{I}}

\newcommand{\rR}{\mathbb{R}}
\newcommand{\TJ}{E}

\newcommand{\one}{\mathbb{I}}
\newcommand{\vq}{\boldsymbol{q}}
\newcommand{\vp}{\boldsymbol{p}}
\def\modif#1{\textcolor{black}{#1}}
\expandafter\let\csname equation*\endcsname\relax 
\expandafter\let\csname endequation*\endcsname\relax 
\usepackage{amsmath}
\usepackage{amssymb}

\begin{document}

\title{Spectral statistics of chaotic many-body systems}

\author{R\'emy Dubertrand$^{1}$\footnote{remy.dubertrand@ulg.ac.be} and Sebastian M\"uller$^{2}$\footnote{sebastian.muller@bristol.ac.uk}}
\address{
$^1$  D\'epartement de Physique, Universit\'e de Li\`ege, 4000 Li\`ege,
Belgium\\
 $^2$ School of Mathematics, University of Bristol, Bristol BS8 1TW, UK}

\begin{abstract}

We derive a trace formula that expresses the level density
of chaotic many-body systems as a smooth term plus a sum over contributions associated to  solutions of the 
nonlinear Schr\"odinger (or Gross-Pitaevski) equation. 
Our formula applies to bosonic systems with discretised positions,
 such as the Bose-Hubbard model,  
 in the semiclassical limit as well as in the limit where the number of particles is taken to infinity. We  use the trace formula  to investigate the spectral statistics of these systems, by studying interference between solutions of the nonlinear Schr\"odinger equation.
We show that in the limits taken the statistics  
of fully chaotic many-particle systems becomes universal and agrees with predictions from the Wigner-Dyson ensembles of random matrix theory. The conditions for Wigner-Dyson statistics involve a gap in the spectrum of the Frobenius-Perron operator, leaving
the possibility of different statistics for systems with weaker chaotic properties.
\end{abstract}

                              
\maketitle


\section{Introduction}

Feynman's path integral (see e.g.~\cite{schulman}) provides a convenient way to represent the propagator of a quantum mechanical system,
and an excellent starting point for semiclassical and related approximations.
Prime examples are van Vleck's approximation of the propagator of a quantum
system as a sum over
contributions of classical trajectories \cite{VanVleck},
and Gutzwiller's seminal work  \cite{Gutzwiller} relating  the energy spectrum of   chaotic single-particle systems to
periodic classical trajectories.
 These semiclassical methods provide one of the foundations of the field of quantum chaos
\cite{Haake,Stoeckmann,Gutzwiller-book}.  For a many-particle system identifying
the semiclassical limit is  less obvious.
A
promising approach is  to consider the path integral in second quantisation, running   over different choices for the macroscopic wave function parameterised by position and time.
One can  show that in the semiclassical limit and in the limit where the number of particles $N$ is taken to infinity this path integral is dominated by stationary points of the action,
corresponding to solutions of the nonlinear Schr\"odinger equation or Gross-Pitaevski equation. These solutions take on a role analogous to the one played by classical
trajectories in the single-particle theory.
However previous work usually focused either on studying the full
problem in second quantisation or, as e.g. in nuclear dynamics \cite{NegeleRMP}, on {\it one} dominating solution to the Gross-Pitaevski equation \cite{BraunGarg,aguiar2}. 
This does not  exhaust the power of the approximation. In particular keeping the sum
over different solutions of the Gross-Pitaevski equation allows to account for crucial interference effects
between such solutions. 
An approach where the  semiclassical propagator of bosonic many-particle systems was used to study these effects  was pioneered in \cite{fock} for  coherent backscattering, see also \cite{fermi} for applications to fermionic
systems. 

In the present paper we will focus on a further fundamental problem for
which the interference between stationary  points of second quantised path integrals is of vital importance: the statistics of the energy
levels of many-body systems. 
To do so we will first derive an
approximation of the level density in terms
of stationary points of the action, and then study
the interference between these points.
To our knowledge the consequences of interference effects for many-body spectral statistics have not yet been investigated explicitly.
We will see that this statistics depends crucially on the  dynamics generated
  by the Gross-Pitaevski equation. If the dynamics is fully chaotic (in
the sense to be specified below) the   statistics in the limits considered
becomes universal, and agrees with predictions from random-matrix theory (RMT).
These predictions entail, for instance, a repulsion between the
energy levels.

This universal behaviour   mirrors the behaviour of chaotic single-particle systems
studied in 
the semiclassical  limit. For such  systems   spectral statistics
faithful to random matrix theory was conjectured in \cite{BGS}, 
 and a semiclassical explanation  was developed   
in \cite{Berry,HOdA,Argaman,SR,TauSmall,TauLarge}.
This explanation is based on Gutzwiller's trace formula \cite{Gutzwiller,Haake,Stoeckmann,BalianBloch}
\begin{equation}\label{trace}
d(E)\approx\bar d(E)+\frac{1}{\pi\hbar}{\rm Re}\sum_p A_p
e^{iS_p/\hbar}\end{equation}
which expresses the level density as a smooth term $\bar d(E)$ plus fluctuations associated
to classical periodic orbits $p$ (with energy $E$) of the system. Here $S_p$ is the reduced
action of the orbit given by
$S_p=\int\boldsymbol{p}(t )\cdot\boldsymbol{\dot
q}(t )dt $
where ${\boldsymbol{q}}(t )$ denotes the vector of generalised coordinates
and ${\boldsymbol{p}}(t)$ denotes the associated momentum. The amplitude $A_p=\frac{T_p^{\rm prim}e^{-i\mu_p\frac{\pi}{2}}}{\sqrt{|\det(M_p-\one)}|}$
depends on the primitive period $T_p^{\rm prim}$, the so-called Maslov index
$\mu_p$
and the stability matrix $M_p$ relating deviations in the end of the orbit
$p$ to deviations in the beginning; $\one$ is a unit matrix. (Throughout this paper
$\one$ will denote unit matrices with a subscript indicating their size if
it is not clear from context.)
Our first aim will be to generalise the trace formula to bosonic many-particle systems in second quantisation, with solutions of the Gross-Pitaevski equation taking the role of classical
trajectories.

We will then use this result to investigate spectral statistics.
An observable characterising spectral statistics is the two-point correlation
function of the level density $d(E)=\sum_j\delta(E-E_j)$ (where $E_j$ are the energy
levels of the system). This correlation function is defined by
\begin{equation}
R(\epsilon)=\frac{1}{\bar d^2}\left\langle d\left(E+\frac{\epsilon}{2\pi\bar
d}\right)d\left(E-\frac{\epsilon}{2\pi\bar
d}\right)\right\rangle
\end{equation}
where    the average $\langle\ldots\rangle$
is taken over
an interval of $E$ for which $\bar d$ can be taken constant as well as over a
small range of $\epsilon$.
Inserting the trace formula one obtains a double sum over solutions of the Gross-Pitaevski equation, and by taking into account interference between
solutions we indeed recover  statistics in agreement with RMT.

More precisely this agreement holds for the statistics inside appropriate
subspectra defined by the symmetries of the problem; for many-body systems we have at least one symmetry, particle number conservation, requiring to consider subspectra associated to a fixed
particle number. Further refinements (to be discussed below) arise
in case of geometrical symmetries. 
 The precise ensemble
to be chosen depends on the behaviour of the system under time reversal. The
most frequent case involves systems invariant under a time-reversal operator
squaring to 1. In this case (assuming that there are no further symmetries)
one has to use   Wigner's and Dyson's Gaussian Orthogonal Ensemble (GOE),
i.e. predictions for spectral statistics are obtained by modelling the Hamiltonian
through a real symmetric matrix, and then averaging over all possible such
matrices with a Gaussian weight. In the absence of time-reversal invariance
one averages instead over the ensemble of Hermitian matrices with a Gaussian
weight, the Gaussian Unitary Ensemble (GUE).

A paradigmatic example for the systems to be considered is the Bose-Hubbard model, a model with $L$ discrete sites (labelled by $k=0\ldots L-1$)
accommodating bosonic particles.
Denoting the creation and annihilation operators for particles at these sites
by $\hat a_k^\dagger$ and $\hat a_k$ the second-quantised Hamiltonian can
be written as \begin{equation}\label{bh}
\hat H=-\frac{J}{2}\sum_k(\hat a^\dagger_{k+1}\hat a_k+\hat a_k^\dagger
\hat a_{k+1})+\frac{U}{2}\sum_k (\hat a_k^\dagger)^2\hat a_k^2
\end{equation}
describing hopping between the sites as well as interaction between particles
on the same site. One can  consider periodic boundary conditions  and set $\hat a_L=\hat a_0$, $\hat a^\dagger_L=\hat a_0^\dagger$.
More generally we are interested in discrete bosonic many-body 
Hamiltonians   that are of the form
\begin{equation}
\hat H=\sum_{kl}h_{kl}\hat a_k^\dagger \hat a_l+\sum_{klmn}U_{klmn}\hat a_k^\dagger
\hat a_l^\dagger \hat a_m \hat a_n
\end{equation}
or have even higher-order interactions; here real
  coefficients imply time-reversal invariance. 

A crucial requirement is  that the underlying `classical
dynamics' is chaotic. This dynamics  is obtained by replacing the creation and annihilation
operators by mutually complex conjugate time dependent  variables $\psi_k^*$,
$\psi_k$ where
$\vpsi=(\psi_0,\psi_1,\ldots)$ can be interpreted as a macroscopic wave function and 
 $\psi_k$ as its value at site $k$.
As $\hat N=\sum_k\hat a_k^\dagger \hat a_k$ is the particle number operator,
the macroscopic wave function is normalised
to have $\sum_k |\psi_k|^2=N$ where $N$ is the particle number.
The associated analogue of Hamilton's equations for $\psi_k$ can be shown
to read 
$
i\hbar\dot\psi_k=\frac{\partial H}{\partial \psi_k^*}
$
and takes the role of a discrete nonlinear Schr\"odinger equation. 

We note that the `classical' Hamiltonian entering into this equation depends
on the ordering
of operators. For the  Bose-Hubbard model the normal-ordered Hamiltonian
in (\ref{bh}) yields an interaction term $\frac{U}{2}\sum_k|\psi_k|^4$;
however if the Hamiltonian is first brought to Weyl-ordered form (with all
possible
orderings of operators in a product contributing symmetrically) before
replacing the operators by macroscopic wave functions one obtains $\frac{U}{2}\sum_{k}(|\psi_k|^4-2|\psi_k|^2+\frac{1}{2})$.

For the Bose-Hubbard model the  dynamics generated
by the discrete nonlinear Schr\"odinger equation has been found to be mainly
chaotic (with chaotic regions of phase space dominating compared to regular ones) in the case of
 several sites and comparable hopping and interaction terms \cite{kol}.
In the same regime, numerical studies suggest spectral statistics in line
with the GOE \cite{buchkol,kollath}; see also \cite{fermions} for fermionic systems.

\modif{To explain the observed faithfulness to RMT, we
follow a semiclassical approach inspired by single-body spectral statistics
as well as \cite{fock}. Our first aim is to derive a trace formula for
second quantised Bose-Hubbard-like systems.
This will be done
in Sect.$~2$, after a brief reminder of the corresponding derivation
for one-body chaotic systems. Special emphasis is placed on the
treatment of conserved quantities.
  In Sect.$~3$ it is demonstrated how the
obtained trace formula can be used in order to predict the spectral statistics
of the system, especially the $2-$point correlation function. We explain
in detail how to generalise the approach previously used for one-body chaotic systems in first quantisation, and we discuss the range
of validity of this approach.
In Sect.$~4$ the consequences of discrete symmetries of
chaotic many-body system, in particular for Bose Hubbard
model are investigated. In Sect.$~5$ we present numerical results supporting our claims.
Some more technical details of the derivation are put in two appendices.}

\section{Trace formula}

\subsection{Trace formula for single-body systems}

In order to prepare our derivation of a  trace formula for bosonic
many-particle systems,
 we want to briefly review the calculation leading to
the trace formula for single-body systems. We refer to
  \cite{Haake,Stoeckmann,Gutzwiller-book,mehlig,Muratore} for   further
 details.

The   level density can be accessed from the trace of the time evolution operator $e^{-\frac{i}{\hbar}\hat
H t}$ via 
\begin{equation}\label{def_d_E}
d(E)={\rm tr}\,\delta(E-\hat H)= \frac{1}{\pi\hbar}{\rm Im}\,i\int_0^\infty dt\,e^{\frac{i}{\hbar}(E+i0)t}\,{\rm tr}\,e^{-\frac{i}{\hbar}\hat
Ht}\ ,
\end{equation}
where an infinitesimal positive imaginary part has been added to the energy to ensure convergence.
The trace of the time evolution operator can itself be expressed as a path integral over   phase space trajectories 
$(\vq(t),\vp(t)) $ 
\begin{equation}\label{trK}
\,{\rm tr}\,e^{-\frac{i}{\hbar}\hat
Ht}=\int D[\vq,\vp]e^{iR[\vq,\vp]/\hbar},
\end{equation}
where  
the action  weighting each   path is determined by the classical Hamiltonian $H$ as\begin{equation}\label{R1}
R[\vq,\vp]=\int_0^t dt' \left({\vp(t')}\cdot\dot{\vq}(t')-H(\vq(t'),\vp(t'))
\right).
\end{equation}
Now we are interested in an approximation of this path integral in the \modif{\emph{semiclassical limit}}, i.e., the limit of large 
\modif{quantum numbers}
implying that typical classical actions are much larger than $\hbar$; formally this limit is often denoted by $\hbar\to 0.$ In the semiclasssical limit the path integral is dominated by stationary points of the action, corresponding to periodic orbits that satisfy  Hamilton's equations of motion, and     a stationary-phase approximation leads to a discrete sum over such periodic orbits. For chaotic
dynamics, free from continuous symmetries, 
the periodic orbits are isolated but one has to take into account that each of them formally gives rise to a one-parameter family of stationary points distinguished by which phase-space point along the orbit is taken as the initial one associated to $t=0$. The Laplace transform in (\ref{def_d_E}) can then be carried out in a further stationary-phase approximation and one obtains the trace formula
\begin{equation}\label{trace_1b}
d(E)\approx\bar d(E)+
\frac{1}{\pi\hbar}{\rm Re}\sum_p \frac{T_p^{\rm
prim}e^{-i\mu_p\frac{\pi}{2}}}{\sqrt{|\det(M_p-\one)}|}e^{iS_p/\hbar}
\end{equation}
already shown earlier. Here the determinant and the   phase factor involving the Maslov index arise from the Hessian matrix of the action entering the stationary-phase approximation, and the factor $T_p^{\rm
prim}$ results from integration over different choices of initial points.
The summand $\bar d(E)$ can be derived from a careful treatment of the lower limit   of the time integral (\ref{def_d_E}).

\subsection{Path integral for many-particle systems} 

We now want to generalise the trace formula to   many-particle systems in second quantisation, for the case of bosonic particles at discrete sites.
In second  quantisation it is natural to work
  in a basis of coherent states, i.e., normalised joint eigenstates of all annihilators $\hat a_k$ with
eigenvalues $\psi_k$. Further differences from
the single-particle setting arise from operator ordering as well as the conservation
of the particle number.

Again we access the level density  from the trace of the time evolution operator $e^{-\frac{i}{\hbar}\hat
H t}$ using (\ref{def_d_E}). 
For many-particle systems the latter trace is given by the path integral
\cite{AS}
\begin{equation}\label{path}
\,{\rm tr}\,e^{-\frac{i}{\hbar}\hat
Ht}=\int D[\vpsi,\vpsi^*]e^{iR[\vpsi,\vpsi^*]/\hbar}
\end{equation}
 over all   macroscopic wave functions $\vpsi(t')$ that return to their initial value after time $t$. Here the action is 
\begin{equation}\label{R1}
R[\vpsi,\vpsi^*]=\int_0^t dt' \left(-\frac{\hbar}{i}\vpsi^*(t')\cdot\dot\vpsi(t')-H(\vpsi^*(t'),\vpsi(t'))
\right).
\end{equation}
 Eq. (\ref{path}) is related to  but somewhat simpler than the path integral for  matrix elements of coherent state time evolution operator \cite{baranger}; our use of ${\rm tr}\,e^{-\frac{i}{\hbar}\hat
Ht}$ is motivated by \cite{mehlig,Muratore}.
 
In case of {\it normal ordering} the path integral can be derived by splitting
the time interval $t$ into ${\cal J}$ steps of width $\tau=\frac{t}{{\cal
J}}$ and then using the result for the short-time propagator 
\begin{equation}\label{matel}
\langle \vpsi_{j+1}|e^{-i\hat H\tau/\hbar}|\vpsi_j\rangle=
\exp\left(\vpsi_{j+1}^*\cdot\vpsi_j-\frac{|\vpsi_{j+1}|^2+|\vpsi_j|^2}{2}-\frac{i}{\hbar}H(\vpsi_{j+1}^*,\vpsi_j)\tau\right)+O(\tau^2);
\end{equation}
after integration over the macroscopic wave functions at all time steps this
leads to a discrete path integral with the action 
\begin{equation}\label{Rdisc}
R[\vpsi,\vpsi^*]= \sum_j \left(-\frac{\hbar}{i}\vpsi^*_{j+1}\cdot(\vpsi_{j+1}-\vpsi_j)-H(\vpsi^*_{j+1},\vpsi_j)\tau
\right)
\end{equation} and then to (\ref{R1}) after taking the limit ${\cal J}\to\infty$.
To fix notation, throughout the paper $j$
will be a time index, $k$ will label degrees of freedom e.g. associated to
sites, and bold vectors assemble all choices for $k$. $j$ is defined modulo
${\cal J}$ and $k$ is taken modulo $L$.
With these conventions the integration measure in the path integral is given by $\prod_{jk}\frac{d\psi_{jk}d\psi^*_{jk}}{2\pi
i}$.

The  integral thus obtained agrees with the phase-space formulation of  the path
integral in first quantisation if the macroscopic wave function is related
to 
 canonical coordinates and momenta
through $\psi_k=\frac{1}{\sqrt{2\hbar}}(q_k+ip_k)$ or $\psi_k=\sqrt{\frac{I_k}{\hbar}}e^{-i\theta_k}$.
We can  now perform a semiclassical approximation valid in the limit $\hbar\to0$.
Here $\psi_k$ is taken of the order $\frac{1}{\sqrt{\hbar}}$
such that the first term in the action (\ref{R1}) becomes independent of $\hbar$; 
\modif{the coefficients in the Hamiltonian are scaled following $J/\hbar\mapsto J$, $U/\hbar^2\mapsto U$ in such a way that the Hamiltonian is independent of $\hbar$ as well}
(We will later also give an interpretation in terms of the limit $N\to\infty$.)
In our limit
the integral is dominated by periodic solutions of the nonlinear Schr\"odinger
equation $i\hbar\dot\psi_k=\frac{\partial H}{\partial \psi_k^*}$ (and its
complex conjugate) following
from the stationarity of $R$.

\subsection{Particle number conservation}
Importantly, however, our Hamiltonian has a continuous (gauge) symmetry w.r.t. multiplying all components of the macroscopic wave function with the same phase factor. A consequence of this symmetry is that as mentioned the total particle number $\hat N$ is a conserved
quantity commuting with the Hamiltonian. It is thus preferable to consider the density of levels forming the subspectrum associated to a fixed particle number.

To implement this restriction we  subject the $I_k,\theta_k$ defined above to a canonical transformation. (See \cite{Creagh,aguiar2} for alternative approaches.) This transformation is chosen to lead  to $I_k',\theta_k'$ where 
$I'_0=\sum_k I_k=\hbar \sum_k|\psi_k|^2=\hbar N$,
the remaining $I_k'$ are linear combinations of the $I_j$, and the $\theta_k'$
are defined such that the overall transformation becomes canonical. If we
choose the transformation in such a way that the range of possible $\theta'_k$
is
limited to $2\pi$ it is  also convenient to
let $\psi'_k=\sqrt{\frac{I_k'}{\hbar}}e^{-i\theta_k'}$.
As $I'_0$ is a conserved quantity   the corresponding canonical coordinate $\theta'_0$ must be absent from the Hamiltonian. The remaining variables
with $k\geq 1$ parameterise
a {\it reduced phase space} associated to the particle number $N$.

We are  interested  in the part of the spectrum associated to a given value
$N$ of the particle number. The associated level density can be formally written as
\begin{equation}\label{getd}
d_{N}(E)={\rm tr}\;\delta_{\hat N,N} {\rm }\,\delta(E-\hat H)=\frac{1}{\pi\hbar}{\rm Im}\,i\int_0^\infty dt\,e^{\frac{i}{\hbar}(E+i0)t}\,{\rm tr}\;\delta_{\hat
N,N}\,e^{-\frac{i}{\hbar}\hat
Ht}
\end{equation}
with the Kronecker delta
\begin{equation}\label{kronecker}
\delta_{\hat N,N}=\frac{1}{2\pi}\int_0^{2\pi}d\phi \;e^{i\phi(\hat N-N)}.
\end{equation}
As one might expect, $d_N(E)$ is determined by solutions periodic
in the reduced phase space. To see this formally we split the exponential $e^{i\phi\hat N}$ from (\ref{kronecker}) into factors 
$e^{i\phi\hat N/{\cal J}}$ to be inserted into each of the short-time propagators
(\ref{matel}).
This gives an additional term
\begin{equation}\label{Rt}
\tilde R[\vpsi,\vpsi^*]=\sum_j\frac{\hbar\phi}{{\cal J}}\psi_{j+1,0}'^*\psi_{j,0}'
\end{equation}
to be added to the action, leading to
\begin{equation}\label{Rtilde}
\tilde R[\vpsi,\vpsi^*]=\int_0^tdt'\frac{\hbar\phi}{t}|\psi'_0(t')|^2
\end{equation}
in the limit ${\cal J}\to\infty$. The phase now becomes stationary if
\begin{equation}
i\hbar\dot\psi'_k=\frac{\partial H}{\partial \psi_k'^*}-\delta_{k0}\frac{\hbar\phi}{t}
\psi_0' \end{equation}
where the addition only affects the dynamics of $\theta'_0$, replacing it
by
\begin{equation}\label{changeddyn}
\dot\theta_0'=\frac{\partial H}{\partial I_0'}-\frac{\phi}{t}.
\end{equation}
Hence, as anticipated, the requirement of periodicity is nontrivial only for the
variables $\psi'_k$ with $k\geq 1$. $I_0'$ is conserved and hence trivially
periodic, and $\theta_0'$ is required to be periodic only after modifying
the dynamics through the $\phi$-dependent term in the action, but all possible
choices of $\phi$ are integrated over. 

\subsection{Determinant of the Hessian matrix}

\label{determinant}

We now have to determine the weight associated to each periodic solution.
We recall that if the stationary points $p$ of a given action $R[\vx]$ (with
$\vx\in\rR^n$ for now) are isolated and we are taking the limit $\hbar\to0$, the integral over $e^{iR/\hbar}$ can be approximated 
by the following sum over contributions associated to stationary
points, 
\begin{equation}\label{sphase}
\int d^nx\;e^{iR[\vx]/\hbar}\approx
\sum_p(2\pi i)^{n/2}\Big|\det\frac{1}{\hbar}\frac{\partial^2 R_p}{\partial\vx^2}\Big|^{-1/2}e^{iR_p/\hbar-i\mu_p\frac{\pi}{2}};
\end{equation}
here $\mu_p$ is the  number of the negative   eigenvalues of the Hessian
matrix $\frac{\partial^2 R_p}{\partial\vx^2}$ for the stationary
point $p$.
If the
stationary points are not isolated this leads to vanishing eigenvalues of the Hessian. In this case the integral over the  directions orthogonal to
the stationary-point manifold can
still be computed using (\ref{sphase}),  but it has to be accompanied by an integral over the manifold itself.
 
The stationary points of (\ref{Rdisc}) are not isolated. In particular continuous time shifts of a solution of the nonlinear Schr\"odinger equation lead to a different
solution, and the same applies to simultaneous shifts of all $\theta'_{j0}$. This is a consequence of the two conserved quantities, for the  energy conjugate to time and the particle number conjugate to the $\theta'_{j0}$.
However as a first step it is still helpful to compute the
  (rescaled) Hessian
involving derivatives w.r.t. all components of the macroscopic wave functions at all time steps. Adopting a complex notation we consider 
\begin{equation}\label{hessian}
\tilde{\cal H}=\frac{1}{\hbar}\frac{\partial^2R}{\partial(\vpsi_0^{*},\vpsi_0,\vpsi_1^{*},\vpsi_1,\ldots)^2}. \end{equation} 
Using the discretised action (\ref{Rdisc}) the derivatives are given by
\begin{align} 
\label{hessian}
 \frac{\partial^2 R}{\partial {\vpsi^2_j}}&=-\frac{\partial^2 H(\vpsi^*_{j+1},\vpsi_j)}{\partial{
\vpsi_j}^2}\tau\nonumber\\
\frac{\partial^2 R}{\partial {\vpsi_j^{*}}^2}&=-\frac{\partial^2 H({\vpsi_{j}^*},\vpsi_{j-1})}{\partial{
\vpsi_{j}^{*}}^2}\tau,\nonumber\\
\frac{\partial^2 R}{\partial \vpsi_{j+1}^{*}\partial \vpsi_j}&=\frac{\hbar}{i}\one-\frac{\partial^2 H(\vpsi^*_{j+1},\vpsi_j)}{\partial\vpsi^{*}_{j+1} \partial\vpsi_j}\tau\nonumber\\
\frac{\partial^2 R}{\partial {\vpsi_{j}^{*}}\partial \vpsi_j}&=-\frac{\hbar}{i}\one
\end{align} 
where $\one=\one_L$ is a unit matrix of dimension $L$. 
The Hessian w.r.t. these variables  
 then assumes periodic block tridiagonal form 
\begin{eqnarray}\label{tridiag}
\tilde{\cal H}=
\begin{pmatrix}A_0& B_0 & & C_{0}\\
C_1&\ddots&\ddots&\\
&\ddots&\ddots&B_{n-2}\\
B_{n-1}&&C_{n-1}&A_{n-1}\end{pmatrix}
\end{eqnarray}
where $n=2{\cal J}$ and the blocks are matrices of size $L\times L$
given by
\begin{align} 
A_{2j}&=-\frac{\partial^2 H(\vpsi_{j}^*,\vpsi_{j})}{\partial{
\vpsi_{j}^{*}}^2}\frac{\tau}{\hbar}+O(\tau^2)\notag\\
A_{2j+1}&=-\frac{\partial^2 H(\vpsi_{j}^*,\vpsi_{j})}{\partial{
\vpsi_{j}}^2}\frac{\tau}{\hbar}+O(\tau^2)\notag\\
B_{2j}&=i\one\notag\\
B_{2j+1}&=-i\one-\frac{\partial^2 H(\vpsi^*_{j},\vpsi_j)}{\partial\vpsi^{ }_{j} \partial\vpsi^{*}_j}\frac{\tau}{\hbar}+O(\tau^2)\notag\\
C_{2j}&=-i\one-\frac{\partial^2 H(\vpsi^*_{j},\vpsi_j)}{\partial\vpsi^{*}_{j}
\partial\vpsi^{ }_j}\frac{\tau}{\hbar}+O(\tau^2) \notag\\
^{ }C_{2j+1}&=i\one .\label{derivs}
\end{align}
Here we neglected corrections of order $\tau^2$ arising from the fact that
the arguments of the Hamiltonian in (\ref{hessian})  are taken at slightly different times.
We can now use a  general formula for determinants of
block tridiagonal matrices  as in (\ref{tridiag}) that was derived in \cite{Matrix}
using a transfer matrix approach,
\begin{equation}\label{det}
\det\tilde {\cal H}=(-1)^{n(L+1)}\det(\tilde M-\one_{2L})\det(P).
\end{equation}
Here we have
\begin{align}
P&=B_{0}\ldots B_{n-1}\notag\\
\tilde M&=\begin{pmatrix}-B_{n-1}^{-1}A_{n-1}&-B_{n-1}^{-1}C_{n-1}\\\one&0\end{pmatrix}
\ldots\begin{pmatrix}-B_0^{-1}A_0&-B_0^{-1}C_0\\\one&0\end{pmatrix}.
\label{det2}
\end{align}
In Eqs. (\ref{det}), (\ref{det2}) we have slightly modified the numbering of indices from \cite{Matrix}
and taken the dimension of the block matrices as $L$. Due to $n=2{\cal J}$
the sign factor in (\ref{det}) is just equal to 1.
To evaluate $\tilde M$ we  group the factors in (\ref{det2}) into pairs
\begin{equation}
\tilde M_j=\begin{pmatrix}-B_{2j+1}^{-1}A_{2j+1}&-B_{2j+1}^{-1}C_{2j+1}\\\one&0\end{pmatrix}\begin{pmatrix}-B_{2j}^{-1}A_{2j}&-B_{2j}^{-1}C_{2j}\\\one&0\end{pmatrix}
\end{equation}
which using (\ref{derivs}) can be simplified to
\begin{equation}
\label{Mj}
\tilde M_j=\one_{2L}+\frac{i\tau}{\hbar}\begin{pmatrix}   \frac{\partial^2H(\vpsi^{*}_j,\vpsi_j)}{\partial{\vpsi^{}_j}\partial{\vpsi^{*}_j}}
&  \frac{\partial^2H(\vpsi^*_j,\vpsi_j)}{\partial
{\vpsi_j}^2}
\\
- \frac{\partial^2H(\vpsi^{*}_j,\vpsi_j)}{\partial{\vpsi^{*}_j}^2}
& - \frac{\partial^2H(\vpsi^{*}_j,\vpsi_j)}{\partial{\vpsi^{*}_j}\partial{\vpsi^{}_j}}
\end{pmatrix}+O(\tau^2).
\end{equation}
One can   now show that multiplication with $\tilde M_j$ maps small deviations $(\delta\vpsi_j^{*},\delta\vpsi_{j}^{})$ from a given solution of the nonlinear Schr\"odinger equation at time $j\tau$ to the resulting deviation at time
$(j+1)\tau$. This follows immediately by linearizing around the equation 
\begin{equation}\vpsi_{j+1}=\vpsi_j+\dot\vpsi_j\tau+O(\tau^2)=\vpsi_j-\frac{\partial
H}{\partial \vpsi_j^{*}}\frac{i\tau}{\hbar}+O(\tau^2)
\end{equation}
and its complex conjugate.
Hence the product 
\begin{equation}\label{M}\tilde M=\tilde M_{{\cal J}-1}\ldots \tilde M_1\tilde
M_0
\end{equation} (understood in the limit ${\cal J}\to\infty$) maps deviations in at time 0 to those at time $t$.

The matrix
$P$ is given by
\begin{align} 
 P&=B_{0}\ldots B_{n-1}=\prod_j\left(1-\frac{i\tau}{\hbar}\frac{\partial^2H(\vpsi^{*}_j,\vpsi_j)}{\partial{\vpsi^{*}_j}\partial{\vpsi^{}_j}}
+O(\tau^2)\right)\notag\\
&=\prod_j\exp\left(-\frac{i\tau}{\hbar}\frac{\partial^2H(\vpsi^{*}_j,\vpsi_j)}{\partial{\vpsi^{*}_j}\partial{\vpsi^{}_j}}
+O(\tau^2)\right);
\end{align}
evaluating its determinant and taking the continuum limit then leads to
\begin{equation}\label{detP}
\det P=\exp\left(-\frac{i}{\hbar}{\rm tr}\int_0^t dt'\frac{\partial^2H(\vpsi ^{*}(t'),\vpsi(t'))}{\partial{\vpsi^{*}}\partial{\vpsi^{}}}\right).
\end{equation}
The factor $(\det P)^{-1/2}$ arising from this term in $(\det\tilde{\cal H})^{-1/2}$
is known as the Solari-Kochetov (SK) phase; it is in line with previous
work about the propagator in normal ordering \cite{solari,kochetov,baranger}. 

\subsection{Treatment of the conserved quantities}

\label{conserved}

We now modify our treatment  to take into account particle number and
energy conservation.

For notational convenience it is helpful to complement our
earlier canonical transformation singling out the particle number by a further
transformation
affecting only the variables with $k\geq 1$. This transformation is defined only in the vicinity of a periodic
solution and leads to $I_1'$ indicating the energy on the orbit, and $\theta_1'$
the time along the orbit. The remaining variables $I_k', \theta_{k'}$ with $k\geq2$
indicate transverse deviations from this orbit.
If desired they can again be turned into complex variables $\psi'_k$ as above\footnote
{This requires suitable rescaling to make both quantities dimensionless and
restrict the range of $\theta'_k$ to $2\pi$.}, and the
vector formed by all $\psi'_k$ with $k\geq 2$ will be denoted by $\vpsi^\perp$. 
The present transformation is analogous to the transformation to parallel and perpendicular
coordinates in the derivation of the trace formula in first quantisation
\cite{Gutzwiller,Haake,Stoeckmann}.
 We note that it cannot be expanded to a transformation in the full phase
space as this would imply integrability.

When determining the weight associated to periodic solutions, it is now crucial
to take into account that   shifting
all time coordinates $\theta'_{j1}$ by the same amount leads to a valid solution,
and the same applies to coordinated changes of the variables $\theta'_{j0}$
conjugate to the particle number. Hence there are two linearly independent ways in
which {\it continuous} changes from one stationary point of the action lead
to a different one. As second derivatives of the action in the associated directions must necessarily be zero the matrix $\tilde {\cal H}$ defined
above must have a two-fold eigenvalue zero. 

To compare this to the behaviour of $\tilde M-\one_{2L}$  we first of all observe that ${\tilde M}$
 maps deviations of the variables
associated to $k=0,1$ only to deviations associated to the same $k$.
Written in terms of $\theta'_k,I'_k$ the stability matrix multiplies deviations
$(\delta \theta'_k,\delta I'_k)$ with
\begin{equation}
{\tilde M}^{(k)}=\begin{pmatrix}\label{mk}
1&b_{k}\\0&1
\end{pmatrix}
\end{equation}
where $b_k=\frac{d\Delta \theta'_k}{dI'_k} $. Here the diagonal elements indicate that the conserved quantities $I_k'$ stay fixed and changes  of $\theta'_k$
are translated into equal changes in the end. In the right upper element
$\Delta \theta'_k$ indicates the increase of $ \theta'_k$
along the orbit, e.g. the period for $k=1$.
 The coefficient $b_{k}$ takes into account that a change of the energy  typically changes the period of the orbit, and similarly a change of the particle number
 typically changes the difference of the initial and final $\theta'_0$.  As a consequence of (\ref{mk}), $\tilde{M}^{(k)}-\one_2=\begin{pmatrix}\label{mblock}
0&b_{k}\\0&0
\end{pmatrix}$ has determinant zero.

To deal with these zeroes one can 
consider a perturbation of the Hamiltonian  \cite{Muratore} that replaces one of the zero eigenvalues of the
Hessian   by a small value $\epsilon$. In the corresponding $\tilde{M}^{(k)}$  a factor $\epsilon{\cal J}$
then enters in the lower left corner, turning the determinant into $-\epsilon{\cal J}
b_k$. A brief account of these perturbative results   for the present case is given in   \ref{apppert}. The determinant of the Hessian matrix ${\cal H}$ omitting the directions associated to conserved quantities
can now be evaluated by considering  perturbations for both $k=0$ and $k=1$ and dividing out the two factors $\epsilon$,
leading to
\begin{equation}\label{detH}
\det {\cal H}= \det(  M-\one_{2L-4})\det(P){\cal J}^2b_{0} b_{1}.
\end{equation}
 
 Here $M$ is defined in analogy to  (\ref{Mj}) and (\ref{M})
but only w.r.t. the variables $\vpsi^\perp$ omitting $k=0,1$.
$M$ maps initial deviations of $\vpsi^{\perp*},{\vpsi^{\perp}}$ to the corresponding final values. This
meaning is precisely equivalent to the stability matrix appearing in the
conventional trace formula.

Our result for $\det{\cal H}$ allows to evaluate (in a 
stationary phase approximation) the integral over all momenta and coordinates apart from
$\theta'_{j0},\theta'_{j1}$, as well as the fluctuations of $\theta'_{j0},\theta'_{j1}$ as $j$ is varied. 

\modif{It remains to consider   the constant (in $j$) Fourier modes of
${\theta'_{j0},\theta'_{j1}}$.
Importantly, if we perform a discrete Fourier transform  of   $\theta'_{jk}$
  and
want the associated Jacobian determinant to be 1, the integration variable
parameterising the constant mode has to be chosen as
$\frac{\sum_j \theta'_{jk}}{\sqrt{\cal J}}$, i.e.
$\sqrt{\cal J}$ times the average of the $\theta'_{jk}$.
More natural parameterisations of the constant Fourier modes, such as by
the average of the $\theta'_{jk}$,
entail a Jacobian $\sqrt{\cal J}$ for each $k=0,1$. These
jointly cancel the ${\cal J}^{-1}$ from $(\det{\cal H})^{-1/2}$.}

\modif{Integration over the constant modes
now leads to multiplication with the integration ranges.
For $\theta'_{0j}$ the range is $2\pi$, and
for the time coordinates the range is normally the period. However if the periodic
solution consists of several repetitions of a shorter `primitive' periodic
solutions, all distinctive stationary points of the action can be accessed
by integration over the period $T_p^{\rm prim}$ of the primitive solution
only. Equating $T_p=T_p^{\rm prim}$ if our solution does not consist
of repetitions
of a shorter one we thus obtain a factor $T_p^{\rm prim}$ in all cases.}

We still have to evaluate the integral over $\phi$
\begin{equation}
\label{phiint}
\frac{1}{2\pi}\int d\phi\; e^{i\tilde R/\hbar-i\phi N}.
\end{equation}
where the $\tilde R$ from (\ref{Rtilde})  equals $\phi I'_0$.
Due to $\frac{d \tilde R}{d\phi}=I'_0$ the stationary phase
condition gives $I_0'=\hbar N$ as expected. Now due to (\ref{changeddyn})
a change of $\phi$ is equivalent
to changing the range  $\Delta\theta'_0$ by an opposite amount. This implies that $\frac{d^2\tilde R}{d\phi^2}=-\frac{d I'_0}{d\Delta\theta'_0}=-b_0^{-1}$.
The $b_0$ thus obtained cancels with the one from (\ref{detH}). 

Altogether the trace of the time evolution operator, restricted to a given value of the particle number, is thus approximated by
the following
sum over
solutions $p$ of the nonlinear Schr\"odinger equation that are periodic with
period $T_{p}=t$  in our reduced phase space
\begin{equation}\label{trprop}
{\rm tr}\;\delta_{\hat
N,N}\,e^{-\frac{i}{\hbar}\hat
Ht}\approx\sum_p \frac{T_p^{\rm prim}e^{iR_p/\hbar-i(\mu_p+\mu_p^{(1)})\frac{\pi}{2}}}{\\ \sqrt{2\pi i\hbar| b_1^{}||\det(M_p-\one)|}}.
\end{equation}
Here   various factors from the integration measure, Eq. (\ref{sphase}),
Eq. (\ref{kronecker}), and the $\theta'_0$ integral have cancelled mutually. 
The result is in line with \cite{mehlig,Muratore}. 
The Maslov index $\mu_p$ counts the  negative eigenvalues of the part of the Hessian matrix ${\cal H}$ associated to $k = 2, 3, \ldots$
(when brought to symmetric form involving derivatives
w.r.t. $I'_{jk}, \theta'_{jk}$). 
The index $\mu_p^{(1)}$ takes a similar role for $k=1$; it is equal to 1 if $b_1<0$ and 0 otherwise. An analogous phase associated to $k=0$ has already been cancelled by the stationary-phase approximation 
of (\ref{phiint}). For notational convenience we also use the Maslov index to absorb  the Solari-Kochetov phase  \cite{solari,kochetov,baranger} 
\begin{equation}
\mu_p^S\frac{\pi}{2}=-\frac{1}{2\hbar}{\rm tr}\int_0^t dt'\frac{\partial^2H(\vpsi ^{*}(t'),\vpsi(t'))}{\partial{\vpsi^{*}}\partial{\vpsi}}.
\end{equation}
We note that like the stability matrix also the action can be taken within our
reduced phase space. The only difference between the two is the $k=0$ contribution to the first term in (\ref{R1}), which is given by 
\begin{equation}
 \int_0^t dt' \left(-\frac{\hbar}{i}{\psi'}_0^*(t')\dot\psi'_0(t')
\right)=\int_0^t dt'I'_0(t')\dot\theta'_0(t')=2\pi \hbar N;
\end{equation}
however for integer $N$ this term has no influence on the phase factor $e^{iR/\hbar}$.
Moreover, as the Hamiltonian is independent of $\theta'_{j0}$, the
change in the dynamics of $\theta'_{j0}$ mentioned above does not affect any
of the quantities entering the trace formula.

\subsection{Level density}

Finally the Laplace transform 
of (\ref{trprop})
\begin{equation}\label{repeat}
d_N(E)={\rm tr}\;\delta_{\hat N,N} {\rm }\,\delta(E-\hat H)= \frac{1}{\pi\hbar}{\rm
Im}\,i\int_0^\infty dt\,e^{\frac{i}{\hbar}(E+i0)t}\,{\rm tr}\;\delta_{\hat
N,N}\,e^{-\frac{i}{\hbar}\hat Ht}
\end{equation}can be performed in a further
stationary-phase
approximation   analogous to \cite{Gutzwiller,Haake,Stoeckmann} (and in fact
similar to the $\phi$ integral above). The phase
in $e^{i(Et+R_p)/\hbar}$ becomes stationary if $E=-\frac{dR_p}{dt}$.
However as shown e.g. in \cite{Haake} $-\frac{dR_p}{dt}$ is precisely the
energy of the solution with period $t$. Hence the result
of the stationary phase approximation will be a sum over all solutions that
are periodic in the
sense above and have fixed energy $E$ rather than fixed period.  
For these
solutions $Et$ then
cancels
with the second term in the action (\ref{R1}),
hence the associated phase will be determined only by the first term $S_p=-\frac{\hbar}{i}\int_0^t
dt' \vpsi^*(t')\cdot\dot\vpsi(t')$ also
referred to as the reduced action. Given that
$\frac{d^{2}R}{d t^2}= -\frac{dE}{d t}=-b_1^{-1}$
the factor $\sqrt{2\pi i\hbar |b_1|}e^{i\mu_p^{(1)}\frac{\pi}{2}}$ arising from the stationary-phase approximation combines nicely with the one from (\ref{trprop}).
Altogether we thus obtain the anticipated
trace formula
\begin{equation}
d_N(E)\approx\bar d_N(E)+\frac{1}{\pi\hbar}{\rm Re}\sum_p \underbrace{\frac{T_p^{\rm
prim}e^{-i\mu_p\frac{\pi}{2}}}{\sqrt{|\det(M_p-\one)}|}}_{=: A_p}e^{iS_p/\hbar}.
\end{equation}The summand $\bar d_N(E)$ gives the smooth part of the level density
and arises from the lower limit 0 of the time
integral. It is  proportional to the
volume of the energy shell in our reduced phase space and given by the pertinent
variant of the Weyl
or Thomas-Fermi formula 
\begin{equation}\bar
d_N(E)=\frac{1}{(2\pi\hbar)^{L-1}}\int dI_1'\ldots dI_{L-1}'d\theta_1'\ldots
d\theta_{L-1}'\delta(E-H(\boldsymbol{I}',\boldsymbol{\theta}'))\,.
\end{equation}
This result (which is more meaningful if one avoids the canonical transformation
in the beginning of subsection \ref{conserved}) is readily obtained   using that for small times no time
slicing is necessary. The action (\ref{Rdisc}) then boils down to $-H(\vpsi^*_{0},\vpsi_0)t$
and the result follows directly after Laplace transformation.
For the semiclassical approach to spectral statistics it will not be
necessary to evaluate the -- often involved -- integral for $\bar d_N(E)$.
As in first quantisation \cite{Hummel} subleading corrections to  this result are to be expected.

\subsection{Discussion}

\modif{Noting that $|\vpsi|^2=N$ the present approximation is valid not only for
$\hbar\to 0$ but also in the limit  $N\to\infty$. In the latter case it is
helpful to rescale $\vpsi\to\sqrt{N}\vpsi$ to avoid changing the normalisation
of the variables in the limit taken. 
If we want the hopping and interaction terms in a Bose-Hubbard like system to
remain comparable we then have to adjust the corresponding
coefficients in a way similar to the case $\hbar\to 0$.
This now entails $J\to J, NU\to U$.
In case of the Bose-Hubbard model the resulting Hamiltonian entering
the trace formula then satisfies
\begin{equation}
H(\vpsi^*,\vpsi)=N \left[-\frac{J}{2}\sum_k(\psi_{k+1}^*\psi_k+\psi_k^*
\psi_{k+1})+\frac{U}{2}\sum_k \psi_k^{*\; 2}\psi_k^2\right]\ ,
\end{equation}
with $U$ and $J$ constant, the normalisation $|\boldsymbol{\psi}|^2=1$
and the large parameter $N$ and the whole action (\ref{Rdisc}) being proportional to
$N$.}
 
If the Hamiltonian is  written in Weyl instead of normal ordering one 
obtains an analogous result however without a Solari-Kochetov phase.
This is in line with \cite{mehlig} as well as results for the propagator in \cite{baranger,aguiar1}.
A derivation along the lines followed here will be given in \ref{appweyl}; it uses the analogue of  the discretised action stated in  \cite{aguiar1}   and  the corresponding Hessian  is again evaluated with the help of \cite{Matrix}.

An alternative derivation of the trace formula  can be based on  
the semiclassical approximation \cite{baranger,aguiar1,aguiar2} of    matrix elements
  $\langle\vpsi^{(f)}|e^{-i\hat Ht/\hbar}|\vpsi^{(i)}\rangle$
in a basis of coherent states. Interestingly,  in this case  $\vpsi(t')$ and $\vpsi^*(t')$ first have to be treated as independent functions
subject to the conditions ${\boldsymbol\psi}(0)={\boldsymbol\psi}^{(i)}$, ${\boldsymbol\psi}^*(t)={{\boldsymbol\psi}^{(f)}}^*$
only. 
The two functions become complex conjugate after evaluating the trace in
a stationary-phase approximation, and the resulting trace formula coincides
with the one obtained above.

\section{Spectral statistics} We are now equipped to study spectral statistics.
Inserting the trace formula into the definition of the two-point correlation function one obtains, as for single-particle systems, the double sum
\begin{equation}\label{dsum}
R(\epsilon)\approx{1}+{\rm Re}\left\langle\sum_{p,p'}
\frac{A_p A_{p'}^*}{(\pi\hbar \bar d)^2}e^{i[{ S_p(E+\frac{\epsilon}{2\pi\bar d} ) -S_{p'}(E-\frac{\epsilon}{2\pi\bar
d})}]/\hbar}\right\rangle.
\end{equation}
The   correlation function is thus expressed in terms of
periodic solutions of the nonlinear Schr\"odinger equation in a way 
that allows to keep track of crucial interference effects. The double sum can be evaluated in the same way as for chaotic single-particle systems.

We  now want to discuss this evaluation in more detail. In doing so we will
 emphasise
the ingredients entering the calculation and check the  conditions under which the reasoning for single-particle systems carries
through to many-particle systems in second quantisation. For the  details of the calculation for single-particle systems based
on these ingredients we refer to the
original literature quoted below  as well as \cite{Haake,Handbook}.

\subsection{Conditions}

The phase space of predominantly chaotic many-particle systems 
typically still has small stability islands. Hence it is important to stress
that
our theory describes  the behaviour of states supported by the chaotic part of phase space. The spectral statistics is dominated by this contribution if the regular parts of phase space
are small in comparison, as in the case of the Bose-Hubbard model in the regimes considered.

In the chaotic part of the phase space our treatment requires
a  gap in the spectrum of the Frobenius-Perron operator.
This condition implies various weaker requirements  such
as ergodicity and hyperbolicity. The Frobenius-Perron operator describes the time evolution of classical phase space densities \cite{Haake},
leading from a density $\rho_0(\vx)$ at time 0 to the density $\rho_t(\vx)=
\int d^n x'\,\delta(\vx-\Phi_t(\vx'))\rho_0(\vx')$ at time $t$. Here $\Phi_t(\vx')$
gives the image of the phase space point $\vx'$ under classical time evolution
over time $t$. The leading eigenvalue  of this operator is 1, with
the associated eigenfunction corresponding to a uniform density on the
energy shell. The remaining eigenvalues can be written as $e^{-\gamma_m t}$
where $\gamma_m$ with ${\rm Re}\,\gamma_m\geq 0$ are the Ruelle-Pollicott resonances. The system is said to have a spectral gap if the remaining ${\rm
Re}\;\gamma_m$
are bounded away from zero so that all modes associated to non-uniform phase
space densities
decay in time with  at least a minimal rate.
For many-particle systems the dynamics required to satisfy this condition is the one induced by the
discrete nonlinear Schr\"odinger equation in the reduced phase space parameterised
by $I'_{k},\theta'_{k}$ ($k=1,2,\ldots,L-1$).

As a further condition we assume for now that our system has no further symmetries
beyond the particle number conservation already taken into account, however
we will extend our treatment to deal with discrete geometric symmetries at a
later point.

\subsection{Diagonal approximation}

When evaluating (\ref{dsum}) it is important to look for pairs of solutions whose (reduced)
actions $S_p,S_{p'}$ are similar as this means that their contributions can interference
constructively. In contrast, terms in (\ref{dsum}) arising from pairs of orbits with large
action differences oscillate rapidly as the energy is varied and are washed out by the energy average.

The simplest pairs of solutions  with similar actions  involve two
solutions that are identical (apart from the slight difference due to the
offset in their energy
arguments). If we neglect the difference of the corresponding factors $A_p,A_{p'}$
and Taylor expand the exponent using that $\frac{dS_p}{dE}$ gives the period $T_p$, the contribution from
identical solutions can be written as the single sum
\begin{equation} \label{diagstart}
R_{\rm diag}(\epsilon)={\rm Re}\left\langle\sum_{p}
\frac{|A_p|^2}{(\pi\hbar \bar d)^2}e^{iT_p\epsilon/\pi\bar d} \right\rangle.
\end{equation}
Under the condition of a spectral gap such sums over periodic orbits can
be evaluated using a sum rule derived in the quantum chaos context by Hannay
and Ozorio de Almeida \cite{HOdA}. In the notation used here it can be written as
\begin{equation}\label{sumrule}
\left\langle\sum_p |A_p|^2\ldots\right\rangle\approx\int_0^\infty \frac{dT}{T}\ldots
\end{equation}
where the dots represent an arbitrary property of the solutions that depends
only on their period $T$.
This rule is a general statistical property of periodic solutions
in systems with a spectral gap; it is very helpful to extract information from these
solutions even in situations where it is  difficult to determine the solutions
individually.
Eq. (\ref{sumrule}) implies that even very long orbits give important collective contributions: while the factors $|A_p|^2$ associated to these orbits decrease with increasing period their number increases, and both effects approximately compensate. Using (\ref{sumrule}) the sum in (\ref{diagstart}) can be evaluated to give
\begin{equation}
R_{\rm diag}(\epsilon)=-\frac{1}{2\epsilon^2}.
\end{equation} 
This result, originally derived in \cite{Berry,HOdA}, is known as the diagonal approximation.
It gives the first nontrivial term in an expansion of the two-point correlation function
in $\frac{1}{\epsilon}$,
after the leading term 1 present in (\ref{dsum}).
For time-reversal invariant systems we also have
to keep track of mutually time reversed solutions, leading to a doubling
of this result.

\subsection{Encounters}

\begin{figure} 
\includegraphics[scale=0.37]{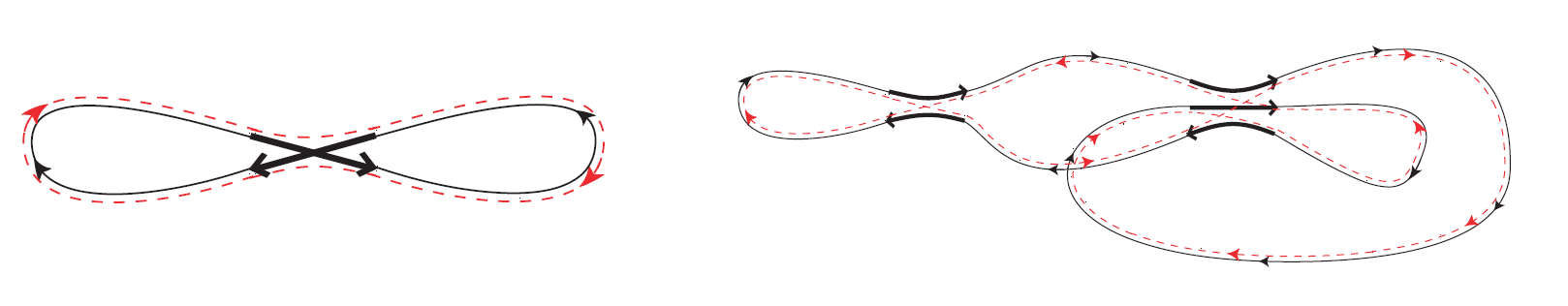}
\caption{Pairs of orbits differing in encounters: Simplified sketch of a
Sieber-Richter pair of orbits differing by their connections in a single encounter of 
two orbit parts, and a pair of orbits differing in two encounters involving
two and three orbit parts.  (Pictures from \cite{TauSmall}\copyright American Physical Society).}
\label{fig:pairs}
\end{figure}

Contributions of higher order in $1/\epsilon$
arise from pairs of orbits differing noticeably only in so-called
encounters \cite{SR,TauSmall}. Inside these encounters two or more parts of the same orbit
come close up to time reversal, and a  partner orbit can  then obtained by
connecting the ends of these orbit parts differently and a subsequent
adjustment to obtain a valid periodic orbit satisfying the equations of motion.
For time-reversal invariant systems one also has to include the case where  parts of
an orbit are almost mutually time reversed. Two simplified sketches of such pairs of orbits  are shown in Fig. \ref{fig:pairs}.

The existence of such pairs of orbits requires hyperbolicity which follows from the existence of a spectral gap. In a hyperbolic system the possible directions at each point in phase space (apart from the direction of the flow and the direction of increasing energy) are spanned by pairs of stable and unstable directions. Deviations between (parts of) trajectories along the stable directions decrease exponentially in time, whereas deviations along stable directions increase exponentially but decrease for large negative times. This allows to `change connections' inside an encounter, by constructing a part of an orbit $p'$ that moves away from one part of $p$  
and towards a different part of $p$;   its deviation from the first part  has to be unstable whereas the deviation from the second part is stable.

When deriving the contribution of such pairs of orbits for single-particle systems,
the deviations between encountering parts of an orbit were measured in a system
of coordinates associated to the stable and unstable directions \cite{TurekRichter,Spehner,Higherdim}.
This carries over in the present scenario as well.
For example, if two parts   come close 
we consider the points where these two parts pierce through a Poincar\'e surface
of section in our reduced phase space. By transformation
from the difference between the  $I'_k$, $\theta'_k$   ($k=2,\ldots,L-1$) we then obtain $L-2$ pairs of  coordinates $s_k,u_k$ characterising
the deviations between the two orbit parts in the stable and unstable directions.
If the parts are almost time-reversed instead of close in phase space
we instead have to consider the deviation of one  part from the time-reversed of the other.

We can now determine the difference between the  (reduced) actions
of the partner orbits. Apart from the difference associated to the energy offset   already seen in the diagonal approximation this has a further
contribution accounting
for the change of action due to the changed connections in the encounters.
For an encounter   of two parts   the latter contribution can be
written as \cite{TurekRichter,Spehner,Higherdim}
\begin{equation}
\Delta S=\sum_k s_k u_k
\end{equation}
where the sum runs over pairs of associated stable and unstable coordinates.
Generalisations to encounters involving more orbit parts are given in \cite{TauSmall}.
As in the phase of (\ref{dsum}) the action difference is divided by $\hbar$, systematic
contributions of constructively interfering orbits will have action differences
of this scale and hence very close encounters.

As a further ingredient one needs to determine the probability that encounters
with given separations between the stretches arise in a periodic orbit/solution.
For the corresponding formula we refer to \cite{TauSmall}.
We stress that the only requirement for its validity is the existence
of  spectral gap. This is used to derive an `ergodic' probability for different parts of an orbit to come close, as well as an estimate for the duration of an encounter, both of which enter the formula.

With all ingredients for the single-particle treatment remaining valid,  the contributions of all `diagrams' of orbits differing in encounters (such
as those displayed in Fig. \ref{fig:pairs}) remain unchanged. These contributions  involve powers
of $1/\epsilon$ with the power increasing for more complex diagrams.
The sum over all  diagrams can be performed using the combinatorial
techniques discussed in \cite{TauSmall}. For systems without time-reversal invariance the contributions   cancel leaving only the diagonal approximation. For time-reversal invariant systems one obtains a series expansion to all orders in $1/\epsilon$. In either case the results agree with the non-oscillatory
terms with coefficients $c_n$ in the random matrix prediction
\begin{eqnarray}\label{RMT}
R(\epsilon)\sim{\rm 1+Re}\sum_{n=2}^\infty(c_n+d_n
e^{2i\epsilon})\left(\frac{1}{\epsilon}\right)^n .
\end{eqnarray}
Here systems without time-reversal invariance (as described by the Gaussian
Unitary Ensemble) have
$c_2=-\frac{1}{2}$ and $d_2=\frac{1}{2}$
and all other coefficients vanish. For time-reversal invariant systems  (as
described by the Gaussian Orthogonal Ensemble) we have $c_2=-1$, $c_n=\frac{(n-3)!(n-1)}{2i^n}$ for $n\geq 3$,
$d_2=d_3=0$ and $d_n=\frac{(n-3)!(n-3)}{2i^n}$ for $n\geq 4$.  We note that the treatment summarised here  assumes that our system (in
the reduced phase space)  
has no further symmetries as these would lead to additional pairs of orbits
with similar or identical actions, such as mutually reflected orbits in a
system with reflection symmetry. The oscillatory terms and  systems with discrete
geometric symmetries will be discussed later.
  
As in case of normal ordering the trace formula is modified to include 
 the  Solari-Kochetov (SK) phase, we have to check that this does not
affect the results arising in the present approach. To do so, we
note that 
 the contributions to the
double sum only involve {\it differences} between phases associated to closeby partner orbits. Due to $\psi\propto\frac{1}{\sqrt\hbar}$ the SK phase itself
is of an order independent of $\hbar$. As the SK phase is time-reversal
invariant the phases of two partner orbits can only differ due to the slight changes inside encounters. However for the encounters relevant for spectral statistics the encounters become very close in the semiclassical limit, meaning that
the difference between the SK phases becomes negligible.

\subsection{Oscillatory contributions}

The random-matrix prediction (\ref{RMT}) for $R(\epsilon)$ also involves
oscillatory contributions proportional to ${\rm Re}\frac{1}{(i\epsilon)^n}e^{2 i\epsilon}$. To access these contributions a more careful semiclassical approximation
is needed \cite{RiemannSiegel,TauLarge}. In this approximation  the level density is
accessed from spectral determinants via
\begin{equation}\label{dfromdet}
d(E)=\frac{1}{\pi}{\rm Im}\frac{\partial}{\partial\eta}\frac{\det(E-\hat H)}{\det(E+\eta-\hat H)}\big|_{\eta=0}.
\end{equation}
 An approximation for the
spectral determinant on the level of the   trace formula is 
\begin{eqnarray}
\det(E-\hat H)&\propto& \exp\left( -\int d E'{\rm tr} (E'-\hat H)^{-1}\right)
\nonumber\\ &\propto& e^{-i\pi\bar N(E)}\sum_\Gamma F_\Gamma(-1)^{n_\Gamma}e^{iS_\Gamma(E)/\hbar}.
\end{eqnarray}
Here the sum is taken over sets of orbits $\Gamma$ with $n_\Gamma$ elements
and cumulative reduced action $S_\Gamma$; the amplitude $F_\Gamma$ depends
on the stability and the Maslov indices of the contributing orbits and $\bar
N(E)$ is the smooth approximation for the number of energy levels below $E$.
However using the spectral determinant allows to incorporate
further quantum mechanical information, in particular the fact that due to
$\hat H$ being Hermitian $\det(E-\hat
H)$ has to be real for real arguments $E$. As shown in \cite{RiemannSiegel} this
leads to an approximation for the spectral determinant where the contributions
from sets of orbits with cumulative periods larger than half of the Heisenberg
tine $T_H=2\pi\hbar\bar d$ are replaced by the complex conjugate of the contributions
from sets with cumulative periods below this threshold. This `Riemann-Siegel
lookalike formula'  
readily generalises to the many-body systems under consideration as its key ingredient, the semiclassical approximation for  
the trace of the resolvent $(E-{\hat H})^{-1}$, can be accessed from the trace of the propagator
as above;  
essentially one only has to omit the restriction to the imaginary part in
(\ref{getd}). 
Again periodicity is   required only in the reduced phase space.  

Using the improved approximation of the level density as well as the same general ideas as outlined above it
is possible to resolve oscillatory contributions as well.
As each level density brings in two determinants, evaluating
the two-point correlation function then requires to study interference between
quadruplets of (possibly empty) sets of orbits. Hence one also needs to take into account
contributions where, say, after changing connections inside encounters an
orbit is broken into two orbits with similar cumulative action. A treatment of these more involved correlations shows that for
chaotic quantum systems $R(\epsilon)$ fully agrees with the predictions from
RMT~\cite{TauLarge}. 

\subsection{Systems without a spectral gap}

Interestingly, there are chaotic systems that do not have a spectral gap but still satisfy some of the ingredients of our calculation, such as hyperbolicity which is required for the existence of     orbit pairs differing in encounters.
For these systems many of the techniques sketched here are applicable, but the final result of Wigner-Dyson statistics does not carry over as e.g. the sum rule (\ref{sumrule}) becomes invalid. An interesting question for   future work is whether   chaotic many-body systems without a spectral gap could be faithful to RMT ensembles that incorporate more information about the problem at hand and reduce to Wigner-Dyson statistics in important
regimes, e.g. the embedded many-body ensembles
\cite{MonFrench,BW} or ensembles   sensitive to the spacial structure of the problem. An example for a situation where the spacial structure is
  important is the continuum limit with diverging number of sites; additional orbit correlations relevant in this case were identified in
\cite{Gutkin}.

\section{Symmetries} 

\label{sec:symmetries}

In a system with additional discrete symmetries one
needs to consider the spectral statistics inside subspectra determined by
the symmetry group. 
A prime example is the discrete (disorder-free) Bose-Hubbard model with periodic
boundary conditions as introduced above. Its symmetry group is the dihedral
group, consisting of the discrete translation $\psi_k\to\psi_{k+1}$ and its iterates; the reflection $\psi_k\to\psi_{L-1-k}$; and combinations of translations and reflection that can be viewed here as reflection about a different centre. 
\modif{More formally the symmetry group for this model is the dihedral group $D_{L}$. Here $D_n$ stands for the dihedral group of order $2n$.}
 The spectrum of the Bose-Hubbard model decomposes into subspectra labelled by the eigenvalues $e^{2\pi i\kappa/L}$ of the discrete translation operator  where $\kappa=0,\dots,L-1$ denotes the quasi-momentum.
The spectra with $\kappa=0$ and (if integer) $\kappa=L/2$ decompose further into components even and odd under reflection, whereas the remaining subspectra come in energy-degenerate pairs $\kappa$, $L-\kappa$ related by reflection.
Altogether we obtain $L+1$ subspectra for odd $L$ and $L+2$ subspectra for even $L$.
A representation-theoretic justification of this decomposition will be given
below and we refer to  \cite{sandvick} for ways to implement the decomposition numerically.

The level density associated to each subspectrum is obtained using a trace formula where the classical orbits, or in the present context the solutions of the
Gross-Pitaevski equation, are required to be periodic (in the sense above)
inside a fundamental domain of the system \cite{Robbins}. (See also \cite{seligmann}.) For the case at hand this domain can be defined e.g. by converting $\psi_k'$ to $\psi_k$ for
a fixed choice of $\psi_0'$ and then imposing certain conditions on $\psi_k$.
Demanding that ${\rm Re}\psi_k$ is smallest for $k=0$ guarantees
that applying the translation operator to a point inside the 
fundamental domain leads to a point outside.
The same applies to most other elements of the symmetry group
but in order to guarantee that reflection about the 0-th site
leads outside the fundamental domain we need an additional requirement such
as ${\rm Re} \psi_1<{\rm Re}\psi_{L-1}$. 

Each subspectrum $\alpha$ (not distinguishing between degenerate subspectra) can  now be associated to an irreducible representation of the symmetry group, and the corresponding trace formula  \cite{Robbins} has a  form similar to (\ref{trace}),
\begin{equation}
\label{symmtrace}
d_\alpha(E)\approx\bar d_\alpha(E)+\frac{1}{\pi\hbar}{\rm Re}\sum_p \chi_\alpha(g_{p}) A_p e^{iS_p/\hbar}\ .
\end{equation}
The only difference from (\ref{trace})
apart from the restriction to the fundamental domain is the additional factor $\chi_\alpha(g_{p})$.
Here $g_p$ is the group element relating the initial and final point of the orbit $p$ if the orbit is considered in the full phase space as opposed to the fundamental domain. 
The character $\chi_\alpha(g_p)$ is the trace of the matrix representing $g_p$ in the representation $\alpha$. As in \cite{Robbins} the derivation of (\ref{symmtrace}) requires that a projection operator on the part
of the Hilbert space associated to our subspectrum is inserted inside the trace taken
over
the Hilbert space; doing this in our many-particle calculations starting
from  (\ref{getd})   leads to exactly the same modifications as observed  in
\cite{Robbins} for single-particle systems.   

To apply (\ref{symmtrace}) to the Bose-Hubbard model we need the irreducible
representations
of the dihedral group \cite{Dihedral}. 
The representation of the translation operator must have an $L$-fold power
$\one$; its two-dimensional representations are therefore be
given by
\begin{equation}
\label{rep1}
\begin{pmatrix}\cos\frac{2\pi\kappa}{L} & -\sin\frac{2\pi\kappa}{L} \\
\sin\frac{2\pi\kappa}{L} & \cos\frac{2\pi\kappa}{L} \\
\end{pmatrix}.
\end{equation}
The reflection operator is then represented by
\begin{equation}
\begin{pmatrix}1 & 0 \\
0 & -1 \\
\end{pmatrix}
\end{equation}
and the representations of all other group elements can be found using that
the matrix representation of a product of symmetry operations must be the
product of the matrix representations.
The general theory   of symmetries in quantum mechanics (see e.g. \cite{Elliot}) implies that the
energy
eigenstates associated to this representation can be grouped into pairs
(representable as vectors) with identical energy. 
One can
show that for the two-dimensional representation at hand this leads precisely to the energy-degenerate subspectra associated
to eigenvalues of the translation operator $e^{2\pi i\kappa/L}$ ($\kappa\neq
0,\frac{L}{2}$) 
as mentioned above.

However if we have $\kappa=0$ or, assuming even $L$,  $\kappa=\frac{L}{2}$ the
matrix (\ref{rep1}) becomes diagonal and the two-dimensional representations
 defined above become reducible. Instead of using these representations the associated subspectra therefore have to
 be described by one-dimensional representations. These represent the translation
 operator either by 1 or in case of even $L$ alternatively by -1, whereas
 the reflection operator can be represented by either $1$ or $-1$ regardless
 of $L$. Again these spectra precisely correspond to what we said above about
the cases $\kappa=
0,\frac{L}{2}$.

In general the consideration of discrete symmetries can change the appropriate RMT ensemble compared to the one expected based on the time-reversal properties alone \cite{Leyvraz}. However using Eq. (\ref{symmtrace}) one can show that this does not happen
for subspectra associated to representations by real matrices \cite{KeatingRobbins,Chris}.
As   the aforementioned representations of the dihedral group are  real the spectral statistics of all its subspectra is thus in line with the GOE as observed numerically in \cite{kollath} as well as in the next section.

\section{Numerical results}

\label{sec:numerics}

To support our results numerically and complement the previous numerical studies   we have performed  a numerical analysis of the chaotic properties of both the quantum and the classical Bose-Hubbard model. Our results support
 Wigner-Dyson statistics as well as mostly classical chaotic dynamics  in the regime where the hopping and interaction terms are comparable.

\begin{figure}[!ht]
    \begin{center}
      \includegraphics[width=0.8\textwidth]{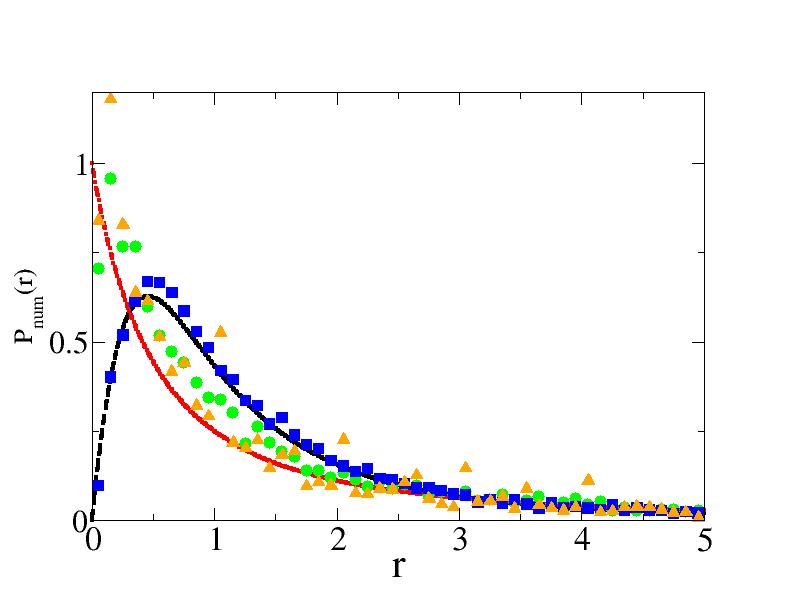}
    \end{center}
\caption{Numerical estimate of the ratio distribution between neighbouring levels for the quantum Bose Hubbard model with $L=5$ and $N=25$.
The dashed black line stands for the RMT prediction (\ref{P_rRMT}). The dotted-dashed red line stands for the Poisson distribution (\ref{P_rint}). Green circles: $UN/J=0.125$. Blue squares: $UN/J=5$. Orange triangles: $UN/J=250$.
} 
\label{examplehist}
\end{figure}

\begin{figure}[!ht]
      
  \begin{minipage}[r]{1.0\linewidth}
    \begin{center}
      \includegraphics[width=0.99\textwidth]{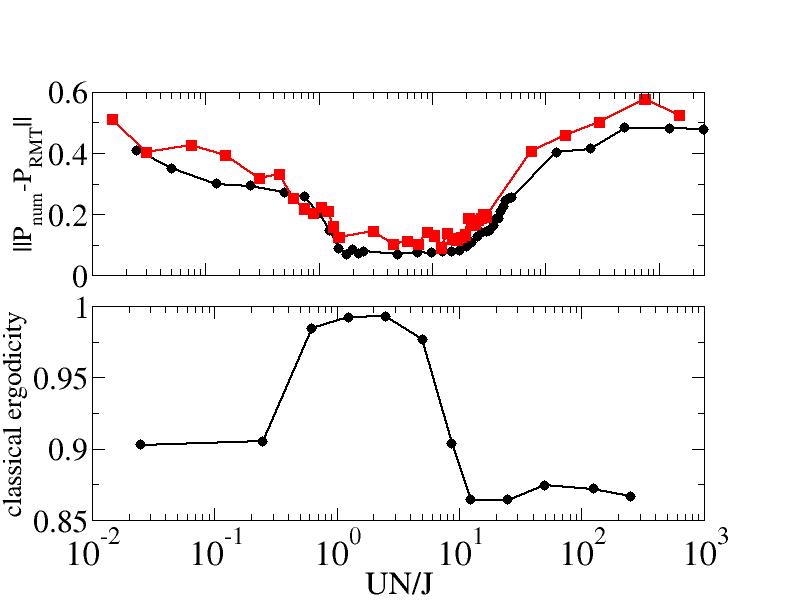}
      
    \end{center}
  \end{minipage}
  \caption{Top: Comparison between the ratio distribution of neighbouring levels between the Bose-Hubbard model and RMT for $N=25$ (black circles) and $N=15$ (red squares), and $L=5$. The difference is estimated via the L$^1$ norm of the deviation from RMT as a function of $UN/J$. Bottom: Numerical 
estimate of
the degree of ergodicity of the classical dynamics of Bose-Hubbard model with $N=25$ and $L=5$ as a function of $UN/J$. $1$ indicates full ergodicity. The precise definition of our estimate is given in the text. 
} 
\label{BHnum}
\end{figure}

For the quantum model we are interested in the spectral statistics as discussed above. We use a statistical observable slightly more convenient for computations than $R(\epsilon)$, namely the normalised distribution $P(r)$ of ratios
between subsequent level spacings; if the ordered quantum levels are denoted
by $E_n$, these  ratios are given by $r_n=\dfrac{E_{n+1}-E_n}{E_n-E_{n-1}}$.
The ratio distribution is especially suited
for our purposes as there is no requirement to explicitly evaluate the
average level density.  A random matrix prediction for $P(r)$ was obtained
in \cite{Atas} by considering
$3\times3$ random matrices; for the GOE it reads 
\begin{equation}
  \label{P_rRMT}
  P_{RMT}(r)=\dfrac{27}{8} \dfrac{r+r^2}{(1+r+r^2)^{5/2}}\ .
\end{equation}
In particular for $r\to0$ one can see that $P(r) \propto r $, which is due
to level repulsion. Notice that for large $r$ the distribution has a fat
tail in contrast to the level spacing distribution: $P(r)\propto r^{-3}$
for $r\gg1$. As for the density of nearest-neighbour spacings the results
for large matrices are expected to be very similar in value but analytically
more complicated.
For comparison, we mention that the ratio distribution for an integrable system (assuming that the energy levels are independent) is given by
\begin{equation}
  \label{P_rint}
  P_{Poisson}(r)=\dfrac{1}{(1+r)^2}\ .
\end{equation}
We determined the spectrum of the quantum Bose-Hubbard model via exact
diagonalisation. For each of the subspectra described in section \ref{sec:symmetries}  we computed the histogram of $r$ to get a numerical estimate of the ratio distribution $P_{\rm num}(r)$. Examples of such numerical histograms are displayed in Fig.~\ref{examplehist}. 
Then we determined the $L^1$ norm (i.e. the integral over the absolute value) of the difference $P_{\rm num}(r)-P_{RMT}(r)$.
Finally this difference was averaged over all different subspectra. 
Fig.~\ref{BHnum}~top shows the norm for the case $L=5$ and $N=15$ as well as $N=25.$  We obtain good agreement for the case that the interaction term
(of order $UN^2$) is comparable to or slightly larger than the
hopping term (of order $JN$). The agreement improves as $N$ increases, which is reassuring as our theoretical derivation of Wigner-Dyson statistics 
holds in the limit $N\to\infty$ (or equivalently in the semiclassical limit).

For the classical Bose-Hubbard model we introduce a numerically determined measure of ergodicity, which indicates whether the classical limit is mostly chaotic.
To  determine  this measure  we considered the Hamiltonian
 \begin{equation}\label{bhclass}
H(\vpsi^*,\vpsi)=-\frac{J}{2}\sum_{k=0}^{L-1}(\psi^*_{k+1}\psi_k + \psi^*_k\psi_{k+1})+ \frac{U}{2}\sum_{k=0}^{L-1} (\psi_k^*)^2\psi_k^2
\end{equation}
where  we identified  $\psi_L=\psi_0$ and $\psi_L^*=\psi_0^*$.
The classical dynamics is given by  \begin{equation}
  \label{class_eq}
  i\hbar\dot\vpsi=\frac{\partial H}{\partial \vpsi^*},\quad i\hbar\dot\vpsi^*=-\frac{\partial H}{\partial \vpsi}.
\end{equation}
If an initial point $(\vpsi^*_0,\vpsi_0)$ is given in phase space, Eq. (\ref{class_eq}) allows to compute the unique trajectory $(\vpsi^*(t),\vpsi(t))$ such that
$(\vpsi^*(0),\vpsi(0))=(\vpsi^*_0,\vpsi_0)$. It may be useful to add that as each point is part of a classical trajectory the point $(\vpsi^*_0,\vpsi_0)$ is parameterised by $2L$ real numbers as the coefficients of $\vpsi^*_0$ are the complex conjugates of the coefficients of $\vpsi_0$.

As the energy  $E$ and the particle number $N$ are conserved, the trajectory can only access a subspace in phase space, denoted by ${\cal M}_{N,E}$, which is the constant energy surface associated to $E$ in the Fock space with
$N$  particles.
Numerically a triangulation of ${\cal M}_{N,E}$ is performed by picking random points on this surface. These points are generated using Halton sequences, which are commonly used to explore high dimensional spaces. First a random point is sampled on the hypersphere $||\vpsi||^2=N$. If its energy coincides with $E$ it is kept and stored as a point in ${\cal M}_{N,E}$. Otherwise we continue the sampling. In order to cover substantially ${\cal M}_{N,E}$ for typical values of $E$, $UN/J$ we chose to pick $2\times 10^6$ random points.
Then the mean distance between any two of these points is calculated, let us call it $\delta$. This gives a typical distance scale for the constant energy surface at a given energy $E$. This typical distance is used to define balls 
of radius $\delta/10$ around each points. 
After having determined a cover of ${\cal M}_{N,E}$ the trajectory starting from $(\vpsi^*_0,\vpsi_0)$ is computed for a sequence of increasing (final) times. 
We took  large enough final values of the propagation time so that the number of visited balls does not vary significantly after this time.

Finally a measure for the ergodicity of a trajectory is defined by the ratio between the number of visited balls and the total number of balls. In order to have generic values, this ergodicity measure is averaged over different choices of initial conditions $(\vpsi^*_0,\vpsi_0)$. Our results are displayed in Fig.~\ref{BHnum}~bottom. By definition our  measure is a real number between $0$ and $1$. $1$ indicates perfect ergodicity. The closer it is to $1$ the more ergodic the trajectory is. 
While not a rigorous check of the conditions for random matrix statistics, our measure indicates whether the system is mostly ergodic or
has
a substantially mixed phase space with large stability islands.  
Fig.~\ref{BHnum}~bottom shows that, for the same range of $UN/J$, the classical dynamics is predominantly ergodic and the quantum system agrees closely with RMT prediction.

Our results are compatible with those obtained through a different approach and a modified version of the model in \cite{cassidy}. There the authors chose a different normalisation of $\vpsi$ and claimed that their version of the classical Bose-Hubbard model becomes more and more classical when increasing $U/J$. This is line with the left part of Fig.~\ref{BHnum}~bottom.

\ack We are grateful to F. Alet, A. Altland, T. Engl, E.-M. Graefe,
A. Lazarides, G. Roux, P. Schlagheck, M. Sieber, J.-D. Urbina and A. Winter for helpful discussions.  A derivation of a trace formula for Bose-Hubbard
models has been obtained independently by T. Engl,  J.-D. Urbina and K. Richter.
We acknowledge financial support from  the EPSRC through grant EP/I014233/1 and the network `Analysis on Graphs and Applications'. R.D. acknowledges financial support from Programme Investissements d'Avenir under the program ANR-11-IDEX-0002-02, reference ANR-10-LABX-0037-NEXT, and from the ARC grant QUANDROPS 12/17-02 and is grateful to the University of Mons for the use of CECI supercomputer.
S.M. is grateful for support from the Leverhulme Trust Research Fellowship
RF-2013-470.

\appendix

\section{Perturbation of the stability matrix and the Hessian}

In this appendix we want to explicitly show how perturbation theory can be used to deal with the vanishing eigenvalues
of the stability matrix and the Hessian considered in subsection   \ref{conserved}, associated to $k=0$ and $k=1$. To avoid some of the complexity
of \cite{Muratore} we specifically consider small perturbations $\delta H$ that depend
only on $\theta'_k$, which does not appear as a parameter of the unperturbed
Hamiltonian. In order to avoid ambiguities we choose $\delta H(\theta'_k)$
periodic in $\theta'_k$ with the period coinciding with the range of values covered by $\theta'_k$. 

We first investigate  how the perturbation modifies the vanishing eigenvalue
of the Hessian ${\cal H  }$. The corresponding eigenvector is associated
to a simultaneous shift of $\theta'_{jk}$ for our $k$ at all time steps $j$.
It is thus convenient to rewrite the Hessian in terms of  derivatives  w.r.t.
$I'_{jk}$, $\theta'_{jk}$  ; this also absorbs the divisor 
$\hbar$ in (\ref{hessian}) and it has the advantage that the Hessian becomes
Hermitian.
The perturbation of the Hessian $\delta{\cal H}$ has
as its only non-vanishing matrix elements the derivatives $-\frac{\partial^2\delta
H(\theta'_{jk})}{\partial{\theta'}_{jk}^2}\tau$. In the same coordinate system
the eigenvector $\boldsymbol{e}$ associated to the vanishing eigenvalue has
identical entries for the ${\cal J}$ components associated to the $\theta'_{jk}$   
associated to our $k$ and the remaining components are zero. First-order perturbation
theory
then yields a perturbed eigenvalue
\begin{equation}
\frac{\boldsymbol{e}\cdot\delta{\cal H}\boldsymbol{e}}
{\boldsymbol{e}\cdot\boldsymbol{e}}=-\frac{1}{{\cal J} }\sum_{j }^{{\cal
 }}\frac{\partial^2\delta
H(\theta'_{jk})}{\partial{\theta'}_{jk}^2}\tau 
\end{equation}
taken as $\epsilon$ in subsection \ref{conserved}.

To study the effect on the stability matrix
we first consider the matrix 
$\tilde M_j^{(k)}$ mapping a deviation of $(\theta'_{jk},I'_{jk})$ to the resulting deviation
after a time step of size $\tau$. We obtain, e.g. by converting Eq. (\ref{Mj}) to the
present system of coordinates,
\begin{equation}
\tilde M_j^{(k)}=\begin{pmatrix}
1+\frac{\partial^2 H(\vtheta_j,\vI_j) }{\partial I'_{jk}\partial\theta'_{jk}}\tau&\frac{\partial^2 H(\vtheta_j,\vI_j)}{\partial {I'}_{jk}^{2}}\tau\\
-\frac{\partial^2 H(\vtheta_j,\vI_j)}{ \partial{\theta'}_{jk}^2}\tau&1-\frac{\partial^2 H(\vtheta_j,\vI_j)}{\partial\theta'_{jk}\partial {I'}_{jk}}\tau
\end{pmatrix}.
\end{equation}
Here the lower left entry vanishes for the unperturbed Hamiltonian but is moved away from zero by the perturbation $\delta H(\theta'_k)$. As a consequence of this perturbation the product given in
(\ref{mk})
\begin{equation}
\tilde M^{(k)}=\begin{pmatrix}
1&\frac{d\Delta \theta_k'}{d I'_k}\\0&1
\end{pmatrix}
\end{equation}
then receives the lower left entry \begin{equation}
 -\sum_j\frac{\partial^2 \delta H(\theta_{jk}')}{\partial {\theta'}_{jk}^2}\tau=
 \epsilon{\cal J}
\end{equation} 
as desired.
In contrast the changes of the other entries do not affect the determinant
to linear order in the perturbation.

\label{apppert}


\section{Weyl ordering}

\label{appweyl}

Finally we want to show explicitly that in line with \cite{baranger,mehlig,Muratore,aguiar2,aguiar1}
the SK phase does not arise in case
of { Weyl ordering}. 
Using the Hamiltonian in Weyl ordering the short-time propagator can be written
as  \cite{aguiar1}

\begin{align}\left\langle\vpsi_j|e^{-\frac{i}{\hbar}\hat H \tau}|
\vpsi_{j-1}\right\rangle\approx2^L\int d[\vw_j,\vw_j^*]
\exp\Big(-2|\vw_j|^2+2\vpsi_j^{*}\cdot\vw_j+2\vpsi_{j-1}\cdot\vw_j^*\nonumber\\-|\vpsi_j|^2/2-|\vpsi_{ j-1}|^2/2-\vpsi_j^*\cdot\vpsi_{j-1}-\frac{i}{\hbar}H(\vw_j^*,\vw_j)\tau\Big)
\end{align}
involving an  integration over a  pair of complex conjugate variables
$\vw_j,\vw_j^*$ that were not required in normal ordering.
The action in the discretised  coherent-state path integral then turns into \cite{aguiar1}
 \begin{eqnarray}
R&=& \sum_j\Big( \frac{\hbar}{i}(-2|\vw_j|^2+2\vpsi_j^*\cdot \vw_j+2\vpsi_{j-1}\cdot
\vw_{j}^{*}
-|\vpsi_j|^2/2-|\vpsi_{j-1}|^2/2\nonumber\\&& 
-\vpsi_{j}^*\cdot \vpsi_{j-1})-H(\vw_j^*,\vw_j)\tau\Big)\label{act}
\end{eqnarray}
and the path integral runs over $\vw_j,\vw_j^*$ as well as $\vpsi_j,\vpsi_j^*$.
The action becomes stationary under variation of  $\vpsi_j,\vpsi_j^*$ if
\begin{equation}
\vw_j=\frac{\vpsi_j+\vpsi_{j-1}}{2}\label{st1}
\end{equation}
and its complex conjugate hold. This 
suggests that in the limit of fine discretisation the $\vw_j$ should be interpreted
as macroscopic wave functions halfway between two discretisation steps. Furthermore
stationarity
w.r.t. variation of $\vw_j,\vw_j^*$  implies\begin{equation}
i\hbar\frac{\vw_j-\vpsi_{j-1}}{\tau/2}=\frac{\partial H(\vw_j^*,\vw_j)}{\partial\vw_j^*}
\end{equation}  
and its complex conjugate, which is the appropriate discretisation of the
nonlinear Schr\"odinger equation $i\hbar\dot\vpsi=\frac{\partial H}{\partial
\vpsi^*}$ for the time interval $\tau/2$. After eliminating $\vw_j$ with the help
of (\ref{st1}) a short calculation shows that the combination of macroscopic
wave functions in the  action (\ref{act})
agrees with the simpler form in (\ref{Rdisc}). 
 
We now have to evaluate the Hessian of $\frac{R}{\hbar}$ and its determinant.
In analogy to subsection \ref{determinant} we start with the Hessian containing all second
derivatives. If we first write the
derivatives w.r.t. all $\vw_j^*,\vw_j$
and then those w.r.t. all $\vpsi_j^*,\vpsi_j$
the Hessian assumes the form
$ i\begin{pmatrix}D & -2E \\
-2E^{T}&E+E^{T} \\
\end{pmatrix}$
where $D$ is block-diagonal with the blocks
\begin{equation}
D_j=2\one_{2L}+\frac{i\tau}{\hbar}\frac{\partial^2H(\vw_j^*,\vw_j)}{\partial(\vw_j^{
*},\vw_j)^2} 
\end{equation}
and the orthogonal matrix $E$ is given by
\begin{equation}
\TJ= \begin{pmatrix} 0 & \one_L &   &   \\
 & \ddots & \ddots &   \\
&  & \ddots & \one_{L}\ \\
\one_{L}\ &  &  & 0 \\
\end{pmatrix}.
\end{equation}
The determinant can  now be evaluated as
\begin{eqnarray*}
&&\det i
\begin{pmatrix}D & -2E \\
-2\TJ^{T} & E+E^{T} \\
\end{pmatrix}\\
&=&\det(2E^T)\det(2E-D(2E^T)^{-1}(-E-E^T))\\
&=& \det(D+D\TJ^2-4\TJ)\\
&=&\det\begin{pmatrix}
D_0+4F&D_0+4F^{T}&&&\\
& D_1+4F & D_1+4F^{T} &&\\
&&\ddots&\ddots &\\
D_{{\cal J}-1}+4F^T&&&& D_{{\cal J}-1}+4F
\end{pmatrix}
\\
& \\
\end{eqnarray*}
with $F= \begin{pmatrix}0&\one_{L}\\0&0\end{pmatrix}$.
Here the factor $i$ dropped out because the dimension of the matrix is a
multiple of 4.
 As the final matrix is of block tridiagonal form it can be simplified  with
the help of (\ref{det}).  
Using  \begin{eqnarray*}
\det(D_j+4F^T)&=&4^{L}+O(\tau^2)\\
(D_j+4F^T)^{-1}(D_j+4F)&=&\tilde M_j+O(\tau^2)
\end{eqnarray*}
(where $\tilde M_j$ is defined as in (\ref{Mj}) but with $\vw_j,\vw_j^*$ taking the
role of $\vpsi_j,\vpsi_j^*$)
and again ignoring terms of quadratic and higher order in $\tau$ in the entries
we obtain \begin{equation} \det
\begin{pmatrix}D & 2\TJ \\
2\TJ^T & -\TJ-\TJ^{T}
\end{pmatrix} =4^{{\cal J}L}\det(\tilde M-\one_{2L}).\end{equation}
As usual the stationary-phase approximation
requires the inverse square root of this determinant, containing a factor
$2^{-{\cal J}L}$. This factor  is cancelled by the $2^{{\cal J}L}$ 
 in the integration measure for $\vw_j,\vw_j^*$. As anticipated, we thus
obtain the same result as with normal ordering apart from the SK
phase.
The remaining steps, in particular taking into account the conservation laws,
carry over directly from the case of normal ordering discussed in the main
text.

\end{document}